\documentclass[usenatbib]{mnras}
\usepackage{ae, amsmath, amssymb}

\title{LIGO Signals from the Mirror World}

\author[Revaz Beradze and Merab Gogberashvili]{
Revaz Beradze,$^{1}$\thanks{E-mail: revazberadze@gmail.com}
and Merab Gogberashvili$^{1,2}$\thanks{E-mail: gogber@gmail.com}
\\
$^{1}$Javakhishvili Tbilisi State University, 3 Chavchavadze Avenue, Tbilisi 0179, Georgia \\
$^{2}$ Andronikashvili Institute of Physics, 6 Tamarashvili Street, Tbilisi 0177, Georgia}

\begin{document}

\maketitle

\begin{abstract}
We want to explain non-observation of electromagnetic counterpart of currently confirmed ten gravitational wave signals from the black hole mergers assuming their existence in the hidden mirror universe. Mirror matter, which interacts with our world only through gravity, is a candidate of dark matter and its density can exceed ordinary matter density five times. Since mirror world is considered to be colder, star formation there started earlier and mirror black holes had more time to pick up the mass and to create more binary systems within the LIGO reachable zone. Totally we estimate a factor 10 amplification of black holes merging rate in mirror world with respect to our world, which is consistent with the LIGO observations.
\end{abstract}



One of the most important scientific achievements of XXI century is the rise of the Gravitational Waves (GW) and multi-messenger astronomy. During the first and second observing runs, the LIGO and VIRGO GW detectors witnessed 10 signals from the binary Black Hole (BH) mergers and one from the binary neutron star system \citep{2018arXiv181112907T}. Using these 10 binary BH GW events data, the merging rate was estimated to be \citep{2018arXiv181112907T}
\begin{equation} \label{LIGO}
\mathcal{R}_{\rm LIGO} = 9.7-101~ \rm Gpc^{-3} \rm yr^{-1} ~.
\end{equation}
From these events source of GW170729 has highest total mass $\sim 80~ M_{\odot}$ and was located further at the luminosity distance $\sim 2.7$~Gpc. The signal GW170608 was emitted by the closest binary BH, with luminosity distance $\sim 300$~Mpc, having the smallest total mass $\sim 20~ M_{\odot}$.

Discoveries made by LIGO indicates that heavy BHs form binary systems and can merge within the age of the universe \citep{2016ApJ...818L..22A}. However, the mechanism of these processes in not fully clear. Most common way for creating a BH is gravitational collapse of a heavy star. In the literature several channels for forming the binary BHs are discussed \citep{2016ApJ...818L..22A}. Isolated massive binary stars can form binary BHs through common-envelope \citep{2018MNRAS.480.2011G}, or via chemically homogeneous evolution \citep{2016MNRAS.458.2634M}. Binary BHs might be formed also in dense stellar clusters by some dynamical processes \citep{2017MNRAS.464L..36A}. Finally, binary BHs can have primordial origin \citep{2018CQGra..35f3001S}.

So called primordial BHs could have been formed from the direct collapse of dark matter density fluctuations in the early universe, when no astrophysical objects existed yet \citep{2018CQGra..35f3001S}. However, relevant for LIGO mass interval of primordial BHs is constrained by several observations. Microlensing surveys restrict number of primordial BHs in the mass range \citep{2001ApJ...550L.169A, 2007A&A...469..387T},
\begin{equation}
10^{-7} ~M_{\odot} < M < 30~ M_{\odot}~.
\end{equation}
The higher mass primordial BHs ($M \gtrsim 43~M_{\odot}$) are excluded by wide binaries \citep{2004ApJ...601..311Y}. The mass range $1-100~ M_{\odot}$ is constrained by the CMB observations, since intense radiation from the ionized gas surrounding primordial BHs, can modify decoupling time of the CMB photons and the ionization history \citep{2008ApJ...680..829R}.

Currently most models assume astrophysical origin of LIGO BHs \citep{2018arXiv181112940T}. These models estimate binary BH merging rate as a function of the efficiency coefficient $\epsilon$, the distribution of times elapsed between creating and merging of a binary system $P$ and BH's birthrate density $\dot N_{\rm BH}$ \citep{2018MNRAS.473.1186E},
\begin{equation} \label{R}
\mathcal{R} = \frac{1}{2}~\epsilon \int P(\tau) \dot N_{\rm BH}~d\tau  ~.
\end{equation}
The dimensionless efficiency coefficient $\epsilon$ depends on many factors and can vary significantly in the interval
\begin{equation}
\epsilon \simeq 0.01-0.001~.
\end{equation}
Delay time $P(\tau)$ is also very speculative as it depends on the binary system configurations and can even exceed the Hubble time. BH birthrate density can be written as \citep{2018MNRAS.473.1186E}
\begin{equation} \label{NBH}
\dot N_{\rm BH} \sim N_{\rm BH}~{\rm SFR(z)} ~,
\end{equation}
where $N_{\rm BH}$ is number density of BHs. Usually it is assumed that $N_{\rm BH}$ is proportional to the total number of stars $N(m)$ in the galaxy of mass $m$,
\begin{equation} \label{N}
N_{\rm BH} \sim N(m)=\frac{m}{\int M\xi(M)~dM}~ ,
\end{equation}
where $\xi(M)$ is a stellar initial mass function (which is integrated in the interval $5~ M_{\odot} < M <150 ~M_{\odot}$ to match the LIGO data).

In (\ref{NBH}) SFR($z$) is a star formation rate, which is typically adopted from the best-fit-function of experimental data \citep{2014ARA&A..52..415M},
\begin{equation} \label{SFR}
{\rm SFR}(z) = 0.015 \frac{(1+z)^{2.7}}{1+[(1+z)/2.9]^{5.6}} ~ \rm M_{\odot} ~ Mpc^{-3} yr^{-1}~.
\end{equation}
This function peaks at $z \sim 1.9$, which corresponds to the lookback time
\begin{equation} \label{t}
t \sim 10.3~{\rm Gyr}~.
\end{equation}

Existing scenarios for explaining the binary BHs merging rate have some lack of confidence. Primordial BHs as LIGO sources are ruled out by CMB and microlensing experiments. Astrophysical BHs in principle can give the merging rate (\ref{R}) near the LIGO's lower bound ($5-10~ \rm Gpc^{-3}yr^{-1}$). However, they require low metallicities of the progenitor stars and certain delay times for binary system merging. Also uncertainties in $P(\tau)$ and $\epsilon$ can vary significantly \citep{2018MNRAS.480.2011G, 2016MNRAS.458.2634M, 2017MNRAS.464L..36A}.

Neutron star merging event was accompanied by the electromagnetic radiation -- $\gamma$-ray burst detected by Fermi \citep{2017ApJ...848L..14G}. While it was unable to witness the electromagnetic counterpart of the GWs from BH mergers \citep{2018arXiv181002764T}. This leads to the idea that all 10 binary BHs were not surrounded by baryonic matter, what seems unnatural.

In this paper we want to suggest a new explanation for LIGO BH events using the mirror world scenario. The binary BH systems that produced GWs could have existed in mirror, or parallel world which interacts with our world only through gravity \citep{1991SvA....35...21K}. Then GWs from binary BHs had no electromagnetic counterparts in our world, because mirror photons cannot interact with ordinary matter. Also mirror matter is a possible candidate of dark matter, and as dark matter is up to 5 times more than the ordinary matter in the Universe, it can increase binary BH merging rate naturally. Unlike primordial BHs, density of mirror black holes is not constrained by the CMB observations. Let us briefly describe the basic concepts of the mirror world model and its cosmological implications, for details see the review \citep{2005ffsc.book.2147B}.

The mirror world was initially introduced to restore a left-right symmetry of the nature. It is assumed that each particle has its mirror partner, which has opposite chirality with respect to its ordinary partner. Mirror particles are invisible for observers from our world and visa-versa. Only possibility for the interaction between these two worlds is gravity and maybe some other unknown weak forces. Alternatively, we can imagine the mirror world scenario as a 5-dimensional theory, with parallel 3D-branes located in two fixed points; Ordinary matter being localized on the left-brane and mirror matter localized on the right-brane, while gravity can freely pass between these two branes.

If mirror sector exists, it was also created by the Big Bang, along with the ordinary matter. But cosmological abundance of ordinary and mirror particle and their cosmological evolution cannot be identical. For instance, the mirror world should have had a lower reheating temperature than our world, which can be achieved in some inflationary models \citep{1996PhLB..375...26B}. If at early times temperatures of two worlds are different and they interact very weakly (through gravity), then they cannot come into thermal equilibrium. Therefore, these worlds will evolve independently during the cosmological evolution and at later stages maintain nearly constant temperature ratio,
\begin{equation} \label{tratio}
x \equiv \frac {T'}{T} < 0.64~,
\end{equation}
which is constrained by the Big Bang nucleosynthesis. In the context of the GUT, or electroweak baryogenesis scenarios, (\ref{tratio}) implies that baryon asymmetry in the mirror world is greater than in our world \citep{2001PhLB..503..362B, 2001PhRvL..87w1304B},
\begin{equation}
1 \leq \frac {n'_b}{n_b}  \lesssim 10~,
\end{equation}
where $n'_b$ and $n_b$ are the number densities of baryons in mirror and our worlds, respectively. If one considers mirror baryon matter as dark matter candidate, this ratio explains near coincidence between densities of visible, $\Omega_b$, and dark (mirror baryon) matter density, $\Omega'_b$, without fine tuning. Also, considering
\begin{equation} \label{x}
x \lesssim 0.2~,
\end{equation}
certain leptogenesis mechanism \citep{2001PhRvL..87w1304B} can imply
\begin{equation} \label{mratio}
\frac {\Omega'_b}{\Omega_b} \approx 5 ~,
\end{equation}
which means that all dark matter can be explained by the mirror matter \citep{2004IJMPA..19.3775B}.

The important moment for structure formation in cosmology is related to the recombination and matter-radiation decoupling epoch, which in ordinary universe happens in the matter domination period at the redshift
\begin{equation}
1+z_{\rm dec} \simeq 1100~.
\end{equation}
But matter-radiation decoupling in the mirror universe can occur earlier \citep{2004IJMPA..19.3775B} and can take place even in the radiation domination era. Indeed, for (\ref{x})
\begin{equation}
1+z'_{\rm dec} \simeq x^{-1} (1+z_{\rm dec}) \simeq 5500~.
\end{equation}

As we have seen, the mirror world evolves alongside our world, with the difference that it has lower temperature and so all the epochs and processes occur earlier. It means that the star formation rate (\ref{SFR}), for (\ref{x}), will peak earlier at $z \sim 9.3$, corresponding to the lookback time in our world
\begin{equation} \label{t'}
t' \sim 13.3~{\rm Gyr}~.
\end{equation}

This implies that mirror BHs have more time to pick up mass and to create binaries in the area covered by the LIGO observations. Compering the lookback times (\ref{t}) and (\ref{t'}) we estimate amplification of merging rate in the mirror world relative to our world by the factor
\begin{equation} \label{A}
A \approx \frac {13.3}{10.3} = 1.3~.
\end{equation}

The additional argument for having more BHs formed in mirror sector is that heavy mirror stars in general evolve much faster than ordinary ones of the same mass \citep{2006APh....24..495B}. Then their way to BH should be faster. Also, after explosion of mirror supernovas, the ejected materials will be reprocessed again and can form new heavy stars.

Besides that, as we mentioned above, the mirror matter density can be 5 times greater than the ordinary matter density (\ref{mratio}). From (\ref{N}) this suggests 5 times bigger star abundance in mirror galaxy and together with (\ref{A}), we derive the factor of $5\times A \approx 6.5$ bigger BH number density (\ref{NBH}) in the mirror world relative to our world. Taking into account the additional argument, associated with the reprocessing of ejected matter, totally we estimate the amplification of binary BH merging rate (\ref{R}) in the mirror world compared to the ordinary world by the factor of order 10. Adopting a typical theoretical value for binary BHs merging rate \citep{2018MNRAS.480.2011G, 2016MNRAS.458.2634M, 2017MNRAS.464L..36A},
\begin{equation}
\mathcal{R}_{\rm theor} \sim 5-10 ~ {\rm Gpc^{-3}yr^{-1}}~,
\end{equation}
we estimate
\begin{equation}
\mathcal{R}_{\rm mirror}  \sim 50-100 ~ {\rm Gpc^{-3}yr^{-1}}.
\end{equation}
So our analysis gives binary BH merging rate density in the upper intervals of LIGO measurements (\ref{LIGO}).

To conclude, inspired by the fact that the 10 currently confirmed GW signals from binary BH systems had no counterpart electromagnetic radiation, in this paper we explore the idea that the sources of these events existed in the hidden mirror universe, which interacts with our world only through gravity. Mirror matter is a candidate of dark matter and its density can exceed ordinary matter density 5 times. Besides that, since the mirror world is considered to be colder, star formation there started earlier and its rate peaks at greater $z$. Totally, we estimated factor of 10 amplification of merging rate in the mirror world with respect to our world. Adopting a common approach, we derived the binary BH merging rate $50-100 ~ \rm Gpc^{-3}yr^{-1}$, which is consistent with the LIGO observations.

\section*{Acknowledgements}

This work was supported by Shota Rustaveli National Science Foundation of Georgia (SRNSFG) [DI-18-335/New Theoretical Models for Dark Matter Exploration].


\bibliography{MNRAS-1}
\end{document}